# Modeling and Controlling Interstate Conflict

Tshilidzi Marwala
School of Electrical and Information
Engineering, University of the Witwatersrand,
P/Bag 3, Wits, 2050 South Africa
E-mail: t.marwala@ ee.wits.ac.za

Monica Lagazio
Department of Political Science
University of the Witwatersrand,
P/Bag 3, Wits, 2050 South Africa
E-mail: lagaziom@social.wits.ac.za

*Abstract*—Bayesian neural networks were used to model the relationship between input parameters, *Democracy*, *Allies*, *Contingency*, *Distance*, *Capability*, *Dependency* and *Major Power*, and the output parameter which is either *peace* or *conflict*. The automatic relevance determination was used to rank the importance of input variables. Control theory approach was used to identify input variables that would give a peaceful outcome. It was found that using all four controllable variables *Democracy*, *Allies*, *Capability* and *Dependency*; or using only *Dependency* or only *Capabilities* avoids all the predicted conflicts.

## I. INTRODUCTION

Recent developments in the liberal peace literature has underlined the importance of treating international conflicts as complex phenomena which displays non-linear patterns of interactions. In this paper, conflict between two states is defined as the threat to use military force or a display of military force, which is conducted in an explicit and overtly non-accidental way [1]. The idea of modeling interstate disputes has challenged the restrictive linear and fixed effect assumptions that have dominated political science. Building on the existing peace theory [2] more interpretations have been suggested and more complex statistical models that are better equipped to deal with the monotonic features of conflict data have been proposed. From the explanation side the relationships between dyadic attributes, which are two states parameters deemed to influence militarized interstates disputes (MIDs), have been interpreted as highly interdependent. MID is defined in this paper as the outcome of interstate interaction and can be either peace or conflict. Beck *et al.* [3] interpret the dyadic attributes as parameters that create a pre-scenario probability of military conflict. This position has been recently confirmed by Lagazio and Russet [4]. Their analysis stresses that low levels of the key variables (economic interdependence, democracy, and shared membership in international organizations) together with distance, relative power, and alliances interact to create multiplicative effects that enhance the likelihood of a dispute, but high levels of those variables do not have the same multiplicative effect on peace. Relative power seems also to exert a strong influence on dispute outcomes when non-democracies are involved but this influence may be much weaker when democracies settle their disputes [5]. Special interstate dependency also needs to be taken into consideration. Having states experiencing conflicts as neighbors may increase the negative influence of some of the dyadic attributes, while having democracies as neighbors may reduce it [6].

In this paper Bayesian neural networks, that are applied using Monte Carlo methods and Gaussian methods [7-9], are used to model the relationships between liberal variables and the MID. Liberal variables are variables such as economic interdependence and democracy, while the MID is 0 for peace or 1 for conflict. Neural networks have been used before to model complex interrelationships between dyadic parameters and the MID [3,4,10]. We also use the Automatic Relevance Determination (ARD) to rank the liberal variables in the order of their influence on the MID [11]. Finally, we introduce a new approach of using traditional control theory to the control of interstate conflict.

## II. MODELING OF CONFLICT

### A. Modeling Data

This section describes the liberal variables and MID data that are used to construct a neural network model. In this analysis we use four variables associated with realist analysis and three "Kantian" variables [12-13]. The first variable is *Allies*, a binary measure coded 1 if the members of a dyad are linked by any form of military alliance 0 in the absence of military alliance. *Contingency* is also binary, coded 1 if both states share a common boundary and 0 if they do not, and *Distance* is the logarithm to the base 10 of the distance in kilometers between the two states' capitals. *Major Power* is a binary variable, coded 1 if either or both states in the dyad is a major power and 0 if neither are super powers. *Capability* is the logarithm to the base 10 of the ratio of the total population plus number of people in urban areas plus industrial energy consumption plus iron and steel production plus number of military personnel in active duty plus military expenditure in dollars in the last 5 years measured on stronger country to weak country. The variable *Democracy* is measured on a scale where 10 is an extreme democracy and -10 is an extreme autocracy and taking the lowest value of the two countries. The variable *Dependency* is measured as the sum of the countries import and export with its partner divided by the Gross Domestic Product of the stronger country. It is a continuous variable measuring the level of economic interdependence (dyadic trade as a portion of a state's gross domestic product) of the less economically dependent state in the dyad. These measures were derived from conceptualizations and measurements conducted by the Correlates of War (COW) project [12-13].

Our data set is the population of politically relevant dyads for the pre-cold war period (PCW), from 1885 to 1945, and the cold war and immediate post-cold war period (CW), from 1946 to 1992, as described extensively and used by Russett and Oneal [2]. For the first population, PCW, only the initial year of the two world wars, 1914 and 1939, is included in the dataset. This restriction insures that the analysis is not unduly influenced by World Wars I and II, and by the absence of adequate trade data for the wartime and immediate postwar years.

We chose the politically relevant population (all dyads containing a major power) because it sets a hard test for prediction. Omitting all distant dyads composed of weak states means we omit much of the influence which variables that are not very amenable to policy intervention (distance and national power) would exert in the full data set. By that omission we make our job harder by reducing the predictive power of such variables, but it also makes it more interesting. By applying the training and validation sampling technique we show that a strong performance is achieved even when the analysis is restricted to the politically relevant group. By focusing only on dyads that either involve major powers or are contiguous, we test the discriminative power of the neural network on a difficult set of cases. The neural network system is fed with only highly informative data since every dyad can be deemed to be at risk of incurring a dispute, yet it is harder for the network to discriminate between the two classes (dyad-years with disputes and those without disputes) because the politically relevant group is more homogeneous (e.g., closer, more inter-dependent) than the all-dyad data set.

The unit of analysis is the dyad-year. There are a total of 27,737 cases in the cold war population, with 26,845 peace dyad-years and 892 conflict dyad-years. The pre-cold war population comprises 11,686 cases, with 11,271 non-dispute dyads and 415 dispute dyads. Only dyads with no dispute or with only the initial year of the militarized conflict are included, since our concern is to predict the onset of a conflict rather than its continuation.

The COW data are used to generate training and testing sets. The validation set is not used because we are pursuing a Bayesian approach to neural network which does not over-fit the model [8-10]. The size of the training sets consists of 500 conflicts and 500 non-conflict data and the test data consists of 392 conflict data and 26345 peace data. The data described in this paper are normalized to fall between 0 and 1. This is because this improves the effectiveness of neural networks modeling [14].

*B. Neural Networks*

In this study, neural networks formulated using Bayesian framework and trained using evidence framework based on Gaussian approximation [8] and Monte Carlo methods [7] were used for fault modeling the relationship between liberal variables and MID. This section gives the over-view of neural networks in the context of a conflict classification problem. In this paper multi-layer perceptron (MLP) [14] supervised learning was used to map the liberal variables ($x$) and the MID ($y$) and this architecture is shown in Figure 1.

The relationship between the $k^{th}$ MID, $y_k$ and the liberal variables, $x$, may be may be written as follows [14]:

$$y_k = f_{outer}\left(\sum_{j=1}^{M} w_{kj}^{(2)} f_{inner}\left(\sum_{i=1}^{d} w_{ji}^{(1)} x_i + w_{j0}^{(1)}\right) + w_{k0}^{(2)}\right) \quad (1)$$

Here, $w_{ji}^{(1)}$ and $w_{ji}^{(2)}$ are first and second layer weights, respectively, going from input $i$ to hidden unit $j$, $M$ is the number of hidden units, $d$ is the number of output units while $w_{j0}^{(1)}$ indicates the bias for the hidden unit $j$, $\omega_{k0}^{(2)}$ indicates the bias for the output unit $k$, $f_{outer}$ is the output layer activation function and $f_{inner}$ is the hidden layer activation function. Selecting the appropriate network architecture is an important part of model building. In this paper, the architecture chosen was the MLP and was trained using the scaled conjugate gradient method [15]. The MLP architecture was chosen using Genetic Algorithms (GA) [16]. The GA used a population of binary-string chromosomes [15,16] and was implemented by performing: (1) simple crossover; (2) binary mutation; (3) and roulette wheel reproduction. Four activation functions were considered and were linear, logistic, hyperbolic tangent and soft-max [14]; and $M$ was restricted to be less than 15. GA population of 20 was used and it identified $M = 10$, logistic function in the output layer and the hyperbolic function in the hidden layers as optimal architecture and converged to a solution after 20 generations.

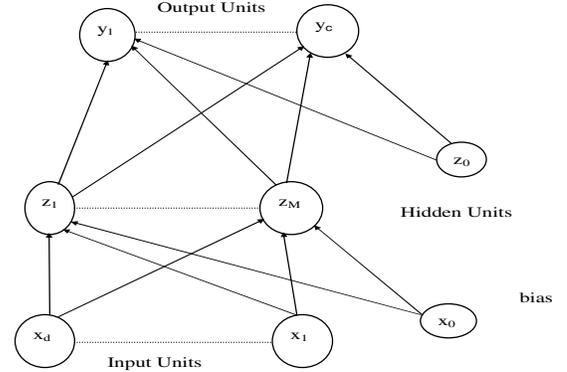

Fig. 1. Feed-forward network having two layers of adaptive weights

The problem of identifying the weights and biases in neural networks may be posed in the Bayesian framework as [8-10]:

$$P(w|D) = \frac{P(D|w)P(w)}{P(D)} \quad (2)$$

where $P(w)$ is the prior distribution function, $D \equiv (y_1,...,y_N)$ is a matrix containing the MID data, $P(w|D)$ is the posterior probability distribution function, $P(D|w)$ is the likelihood function and $P(D)$ is the normalization function known as the "evidence". For the MLP equation 2 may be expanded using the cross-entropy error function to give [14]:

$$P(w|D) = \frac{1}{Z_s} exp\left(\beta \sum_{n}^{N}\sum_{k}^{K}\{t_{nk} ln(y_{nk}) + (1-t_{nk})ln(1-y_{nk})\} - \sum_{j}^{W}\frac{\alpha_j}{2}w_j^2\right) \quad (3)$$

where

$$Z_s(\alpha,\beta) = \left(\frac{2\pi}{\beta}\right)^{N/2} + \left(\frac{2\pi}{\alpha}\right)^{W/2} \quad (4)$$

The cost-entropy function was used because of its classification advantages [14] and the weight-decay for the prior distribution was assumed because it penalises weights of large magnitudes. In equation 3, $n$ is the index for the training pattern, hyperparameter $\beta$ is the data contribution to the error, $k$ is the index for the output units, $t_{nk}$ is the target output corresponding to the $n^{th}$ training pattern and $k^{th}$ output unit and $y_{nk}$ is the corresponding predicted output. The parameter $\alpha_j$ is hyperparameter and it determines the relative contribution of the regularisation term on the training error. In equation 3 the hyperparameters may be set for groups of weights. Equation 3 can be solved in two ways: by using Taylor expansion and approximating it by a Gaussian distribution and applying the evidence framework [9] or by numerically sampling the posterior probability using Monte Carlo method [8]. In this paper both approaches are pursued and the two formulations are compared in the context of the conflict modelling problem.

### C. Gaussian Approximation

Bayesian training of MLP neural networks is essentially about the calculation of the distribution indicated in equation 3. One method of achieving this goal is to assume a Gaussian approximation of the posterior probability by Taylor expansion. If this assumption is made, the posterior probability of the MIDs can be calculated by maximizing the evidence [8]. The evidence is the denominator in equation 2. The evidence framework finds the values of the hyperparameters that are most likely, and then integrates them over the weights using an approximation around the probable weights and the resulting evidence is maximized over the hyperparameters. The evidence framework is implemented by following these steps: (1) Infer the parameters $w$ for a given value of $\alpha$. This is calculated in this paper by using the scaled conjugate gradient optimization method [15]; (2) Infer that the value of $\alpha$ by approximating equation 3 with a Gaussian distribution and maximizing the evidence given the most likely weights.

### D. Hybrid Monte Carlo (HMC) method

Another way of sampling the posterior distribution in equation 3 is the HMC method [17], which is exact, provided that the number of samples approaches infinity. The HMC method uses the gradient of the neural network error to ensure that the simulation samples throughout regions of higher probabilities. This causes the HMC to avoid the random walk associated with traditional Monte Carlo methods. The details of the HMC are in Neal [7]. As a result the HMC performs better than the traditional Monte Carlo method. The gradient is calculated using the backpropagation method [14]. Sampling using the HMC is conducted by taking a series of trajectories and either accepting or rejecting a resulting state at the end of each trajectory. Each state is represented by the network weights and its associated momentum, $p_i$. Each trajectory is achieved by following a series of leapfrog steps which are described in detail by Neal [7]. For a given leapfrog step size, $\varepsilon_0$, and the number of leapfrog steps, $L$, the dynamic transition between two states of the HMC procedure is conducted as follows: (1) Randomly choose the direction of the trajectory, $\lambda$, to be either –1 for backward trajectory and +1 for forward trajectory; (2) Starting from the initial state, $(w,p)$, perform $L$ leapfrog steps with the step size $\varepsilon = \lambda\varepsilon_0(1+0.1k)$, resulting in state $(w_{new},p_{new})$. Here $p$ is the momentum vector and is described in detail in Neal [7], $\varepsilon_0$ is a chosen fixed step size and $k$ is the number chosen from a uniform distribution and lies between 0 and 1; and (3) Reject or accept $(w_{new},p_{new})$ using the Metropolis criterion [18]. In the Metropolis criterion, if the current posterior probability given the weights and the data is higher than the previous posterior probability then accept the new sample otherwise accept it with a low probability.

### E. Neural Network Results and Model Interpretation

Neural networks methods were implemented and the classification of conflict results obtained. To assess the performance of the classification results, the Receiver Operating Characteristics (ROC) graphs are used [19] and are shown in Figure 2. In ROC curve the x-axis is the false positive rate and the y-axis is a true positive rate. The area under the ROC curve indicates how good the classifier is. If the area under the ROC curve is 1 then the classifier has classified all cases correctly while if it is 0.5 then the classifier has classified all cases incorrectly. In the ROC curve in Figure 2, both the classifiers give an area under a ROC curve of 0.82 which is a good classification rate. These results indicate that the Gaussian approximation to the posterior probability and the HMC approach give the same level of classification accuracy.

When a confusion matrix was used to analyze the classification results of the two Bayesian methods, the results in Table 1 were obtained. The confusion matrix contains information about actual and predicted classifications done by a classification system. When the accuracy of the methods used in this paper was calculated by finding the percentage of fault cases that were classified correctly, it was found that the two approaches give results that are statistically similar (both methods give the accuracy of approximately 74% even though the HMC gives marginally more accurate results). A true positive rate is defined as the proportion of positive cases that are correctly identified. The results in Table 1 give a true positive rate of approximately 71% for Gaussian approximation and 73% for the HMC method. A true negative rate is defined as the proportion of negative cases that are classified correctly. The true negative rate calculated from Table 1 indicates that both methods give a rate of 74% even though the HMC performed marginally better than the Gaussian approximation.

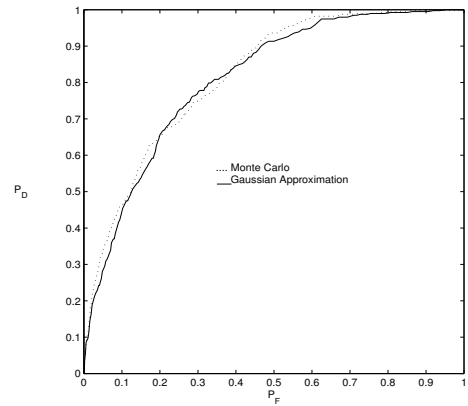

Fig. 2. The ROC of the classification of militarized interstate disputes

However, the HMC was found to be marginally more accurate than the Gaussian approximation. The results in this table show that the HMC is marginally more accurate than the Gaussian approximation. This is primarily due to the fact that the Gaussian approximation is generally not as valid as the Monte Carlo approach [14].

We now interpret the causal model, given by the neural networks developed in this paper. The interpretation of the

TABLE I
CLASSIFICATION RESULTS

| Method | TC | FP | TP | FC |
|---|---|---|---|---|
| Gaussian Approximation | 278 | 114 | 19462 | 6883 |
| Hybrid Monte Carlo | 286 | 106 | 19494 | 6851 |

TC=true conflict (true positive), FC=false conflict (false positive), TP=true peace (true negative), and FP=false peace (false negative)

causal hypotheses represented by a trained neural network is a complex exercise for several reasons. First, neural network models encode their knowledge across hundreds or thousands of parameters (weights) in a distributed manner. These parameters embed the relationships between the input variables and the dependent output. The sheer number of parameters and their distributed structure make the task of extracting knowledge from the network a difficult one. Second, the weight parameters of a multi-layer network usually represent non-linear and non-monotonic relationships across the variables, making it difficult to understand both the relative contributions of each single variable and their dependencies.

When analyzing the causal relationships between input and output variables, the neural network shows that when the dyadic *Democracy* variable is increased from a minimum to a maximum, while the remaining variables are set to a minimum, then the outcome moves from conflict to peace. When all the variables were set to a maximum then the outcome was peace. When all the parameters were set to a minimum then the possibility of war was 52%. When one of the variables was set to a minimum and the rest set to a maximum, then it was observed that the outcome was also always peace. When each variable was set to a maximum and the remaining variables set to a minimum then the outcome was peace except when the variable was either *Contingency* or *Major Power*.

*F. Influence of Model Parameters using the Automatic Relevance Determination (ARD)*

This section introduces and implements the ARD to understand the influence of the input parameters on the MID. The ARD model [9, 11] is a Bayesian model that is used to determine the relevance of each input on the output. The ARD is constructed by assigning a different hyperparameter to each input variable and estimating the hyperparameters using a Bayesian framework. The input weights that have higher hyperparameters are not influential and have less effect on the output than the lower hyperparameters.

In this paper the ARD is used to rank liberal variables with regards to their influence on the MID. The ARD was implemented, the hyperparameters calculated and then the inverse of the hyperparameters was calculated and the results are shown in Figure 3. Figure 3 indicates that the *Dependency* variable has the highest influence, followed by *Capability*, followed by *Democracy* and then *Allies*. The remaining three variables, i.e. *Contingency, Distance* and *Major Power,* have similar impact although it is smaller in comparison with other four liberal variables. The results in Figure 3 indicate that all the liberal variables used in this paper influence the conflict and peace outcome. Thus the proximity, alliance, and power play a part in providing opportunities and incentives for interstate action and therefore have some effects on the peace or conflict between states. Overall, the results, first, supports the theory of democratic peace which claims that democracies never go to war [5]. In Figure 3 it is clear that *Democracy* is a major factor on interstate peace and conflict. Second, the liberal peace theory that economic interdependence promotes peace is demonstrated in Figure 3 by the liberal parameter *Dependency* [20]. However, three variables, *Logdistance, Contingency*, and *Major Power*, cannot be ignored. For example the distance between countries is an important variable with regards interstate disputes. For example, Swaziland and Bahamas have a lower probability of going to wars, primarily, because they are so far apart. However, the influence of *Major Power* cannot be ignored because powerful countries have the capacity to engage in distant conflicts. In summary, the relationship of democracy and interdependence and interstate conflicts is to some extent mediated by both the dyadic balance of power and geographical proximity.

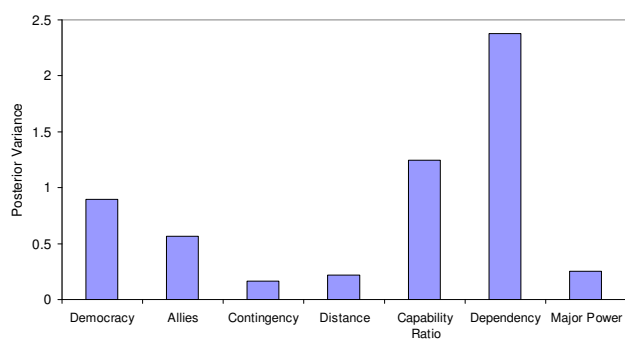

Fig. 3. A graph showing the relevance of each liberal variable with regards to the classification of MIDs.

## III. CONTROL OF CONFLICT

Now that we have developed a model that predicts the MID given liberal variables, the next step is to use this model to identify a set of liberal variables that ensure that the outcome is the desired one. The whole rationale behind the development of the interstate dispute prediction model is to maximize the occurrence of peace. This is achieved in this paper by applying control theory to conflict resolution. Control theory has been used to control many complex problems. A literature review on the application of control system, to solving complex problems, can be found in [21]. This paper reviews recent developments of bioprocess engineering that include the monitoring of the product formation processes. It also reviews the advanced control of indirectly evaluated process variables by means of state estimation using structured and hybrid models, expert systems and pattern recognition for process optimization. Control system theory has been applied to aerospace engineering to actively control the pressure oscillations in combustion chambers [22]. Genetic algorithms and fuzzy logic have been successfully used to control the load frequency in PI controllers [23]. Plant growth has been optimally controlled using neural networks and genetic algorithms [24] and fuzzy control has been used for active management of queuing problem [25]. Other applications of applications of control methods to complex systems may be found in [26-28]. In this paper, we use control system theory for the first time to control interstate conflict. This is conducted by identifying controllable liberal variables that will give a peaceful outcome. To achieve this, the cost function is defined as the absolute value of the neural network prediction, which should be as close as possible to zero, i.e. absolute peace. Two approaches are used and these are: a single strategy approach where only one controllable liberal variable is used and a multiple strategy where all the controllable variables are used. Of the 7 liberal variables discussed earlier in the paper, there are only 4 that are controllable and these are: *Democracy*, *Allies*, *Capability* and *Dependency*.

In this paper, the control system model consists of three components. These are: the feed-forward neural network that takes the liberal variables and predict the MID as well as the optimizer which is activated only if the predicted outcome is war, and therefore undesirable, and its function is to identify the controllable input parameters that predict peace. The approach is illustrated in Figure 4. The optimizer can be any nonlinear function minimization method. In this study the Golden Section Search (GSS) method [29,30] was used for single strategy and Simulated Annealing (SA) [31] was used for the multiple strategy approach. The use of the GSS method is primarily because of its computational efficiency. It should be noted here that other methods such as the conjugate gradient method, scaled conjugate methods or genetic algorithm may also be used to give the same results [29].

On implementing the control strategies the Bayesian neural networks that implement HMC for training was used. The control approach was implemented to achieve peace for the conflict data in the test set. There are 392 conflict outcomes in the test set of which 286 were classified correctly using the HMC trained neural networks (see Table 1). Therefore, in this paper we control 286 conflict outcomes by identifying the controllable liberal variables that will give peace. When the control strategies were implemented, the results shown in Figure 5 were obtained. These results show that for a single strategy approach, where *Democracy* is the controlling variable, 90% of the 286 conflicts could have been avoided. When the controlling variable *Allies* is the only controlling variable used to bring about peace it was found that 77% of 286 conflicts could have been avoided. When *Capability* was used as a controlling variable, 100% of 286 conflicts could have been avoided. When the *Dependency* variable was used as a controlling variable, it was found that all 286 conflicts could have been avoided. When all the controllable variables were used simultaneously to bring about peace, all 286 conflicts were avoided.

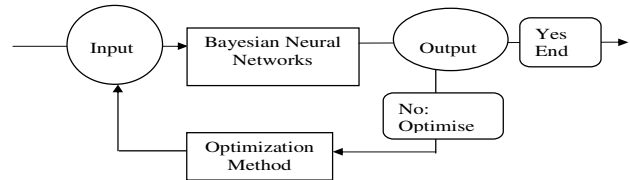

Fig. 4. Feedback control loop that uses Bayesian neural networks and an optimization method

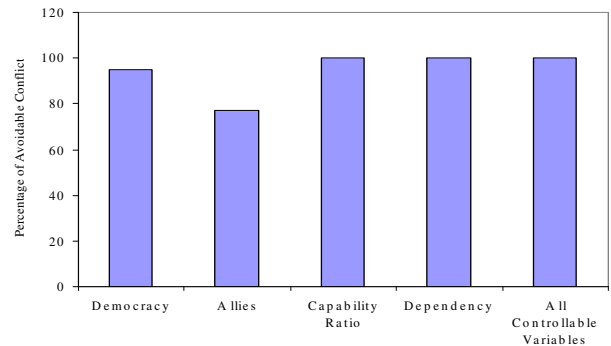

Fig. 5. Graph showing the proportion of past conflicts that could have been avoided.

The results showing the liberal variables that gave conflict outcome, and the liberal variables obtained using a single strategy and a multiple strategy methods to obtain a peaceful outcome are shown in Figure 6. This figure shows that both approaches can give the same outcome. Therefore, the user of the proposed method should decide which approach is the most convenient with regards to the ease of use. Now that we have discussed the controllable liberal variables and the results of various control strategies, we now discuss how these variables can be controlled. There are various ways in which *Democracy* can be controlled and these include linking aid with democracy, pressure of the public opinion, support for opposition groups and embargoes. The variable *Allies* can be controlled by making alliances easier to enter, making them easier to implement and profitable.

The variable *Capability* can be controlled by increasing military expenditure and militarization. The variable *Dependency* can be controlled by increasing trade and economic links. It is observed, in this paper, that an efficient approach to increasing peace in the world is to increase *Democracy*, *Dependency, Capability* and *Alliances* in countries. We, therefore, recommend to all policy makers that the process of increasing *Democracy*, *Dependency*, *Capability* and *Alliances* be put in practice to promote world peace.

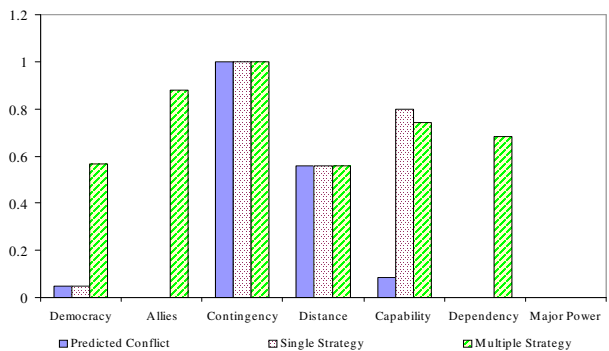

Fig. 6. Results showing the original liberal variables that gave conflict outcome, the results from single strategy approach that gave a peace outcome and the results from a multiple strategy approach that gave a peace outcome

## IV. Conclusion

In this paper Bayesian neural networks were used to model the relationships between input variables, *Democracy*, *Allies*, *Contingency*, *Distance*, *Capability*, *Dependency* as well as *Major Power* and output which may either be *peace* or *conflict*. Gaussian approximation to the posterior probability and hybrid Monte Carlo (HMC) were used to train the neural networks and it was found that the HMC is marginally more accurate than the Gaussian approximation. The automatic relevance determination was used to rank the importance of each input variable and it was found that *Dependency* carry the most weight, then *Capability*, then *Democracy* and then *Allies*. Two control approaches were used to identify input variables that give peaceful outcome. The single strategy approach was implemented using golden section search method and simulated annealing was used for the multiple strategy approach. It was observed that using all four controllable liberal variables simultaneously or only using *Dependency* or using only *Capability* avoids all the previously predicted conflicts; followed by using only *Democracy* (90%) and then *Allies* (77%).